# Dependence of the Efficiency of Spin Hall Torque on the Transparency of Pt-Ferromagnetic Layer Interfaces

Chi-Feng Pai[1], Yongxi Ou[1], D. C. Ralph[1,2], and R. A. Buhrman[1,*]

[1]Cornell University, Ithaca, New York 14853, USA

[2]Kavli Institute at Cornell, Ithaca, New York 14853, USA

We report that spin current transport across Pt-ferromagnet (FM) interfaces is strongly dependent on the type and the thickness of the FM layer and on post-deposition processing protocols. By employing both harmonic voltage measurements and spin-torque ferromagnetic resonance measurements, we find that the efficiency of the Pt spin Hall effect in exerting a damping-like spin torque on the FM ranges from < 0.05 to > 0.10 under different interfacial conditions. We also show that the temperature dependence of the spin torque efficiencies for both the damping-like torque and field-like torque is dependent upon the details of the Pt-FM interface. The "internal" spin Hall angle of the Pt thin films used in this study, after taking the interfacial spin transmission factor into account, is estimated to be ~ 0.20. This suggests that a careful engineering of Pt-FM interfaces can improve the spin-Hall-torque efficiency of Pt-based spintronic devices.

The spin Hall effect (SHE) [1, 2] causes an electrical current density $J_e$ flowing through a material with strong spin-orbit interactions to generate a transverse spin current density $J_s$. The amplitude of $J_s$ is characterized by the spin Hall angle $\theta_{SH} \equiv (2e/\hbar) J_s / J_e$. The most straightforward way to determine a lower bound [3], $\theta_{SH}^{LB}$, on the spin Hall angle in normal metal (NM) systems is to measure the current-dependent torque that is exerted on an adjacent ferromagnet (FM) when spin current flows to the NM-FM interface. Research has shown [4, 5] that there are two different components of torque that can be observed in this case: a "damping-like" torque $\vec{\tau}_{DL} \propto J_e \hat{m} \times (\hat{\sigma} \times \hat{m})$, where $\hat{m}$ is the orientation of the ferromagnetic moment, and a "field-like" torque $\vec{\tau}_{FL} \propto -J_e \hat{\sigma} \times \hat{m}$. The determination via $\tau_{DL}$ measurements of a large $\theta_{SH}^{LB} \approx$ 0.07, in Pt-FM thin film bilayers [3, 6, 7], and the subsequent observation of even larger, "giant" spin Hall angles for high resistivity Ta (amorphous or β phase Ta), $|\theta_{SH}^{LB}| \approx 0.12$ [8] and β-W, $|\theta_{SH}^{LB}| \approx 0.33$ [9, 10], have opened up a very active area for research and technology development.

Recent calculations utilizing Boltzmann analysis [11] and the drift-diffusion approximation [12] have noted that if the NM-FM interface is not completely transparent to the flow of the spin current then spin backflow will reduce the torque applied to the FM by the SHE. This reduction can be characterized, as suggested above, by defining a damping-like spin torque efficiency $\xi_{DL}$, and also a field-like torque efficiency $\xi_{FL}$, for a particular NM-FM interface, such that $\xi_{DL}$ can be less than or equal to the "internal" spin Hall angle $\theta_{SH}$ that quantifies the spin current generated in the absence of an adjacent FM. Within a diffusive model, the effects of spin backflow are expected to modify the spin torque efficiencies in the form [11, 12]

$$\xi_{DL} = \theta_{SH} \times \text{Re} \frac{2G^{\uparrow\downarrow} \tanh\left(\frac{d_{NM}}{2\lambda_{s,NM}}\right)}{\frac{\sigma_{NM}}{\lambda_{s,NM}} + 2G^{\uparrow\downarrow} \coth\left(\frac{d_{NM}}{\lambda_{s,NM}}\right)}, \text{ and } \xi_{FL} = \theta_{SH} \times \text{Im} \frac{2G^{\uparrow\downarrow} \tanh\left(\frac{d_{NM}}{2\lambda_{s,NM}}\right)}{\frac{\sigma_{NM}}{\lambda_{s,NM}} + 2G^{\uparrow\downarrow} \coth\left(\frac{d_{NM}}{\lambda_{s,NM}}\right)}, \quad (1)$$

where $G^{\uparrow\downarrow}$ is the spin-mixing conductance of the interface [13, 14], $G_{NM} \equiv \sigma_{NM} / \lambda_{s,NM}$ is the spin conductance of the NM (where $\sigma_{NM}$ and $\lambda_{s,NM}$ represent the conductivity and the spin diffusion length, respectively), and $d_{NM}$ is the NM layer thickness. If we use previously-determined values for the parameters: $\xi_{DL} \approx 0.07$ (for $d_{Pt} >> \lambda_{s,Pt}$), $\lambda_{s,Pt} \approx 1.4$ nm [3], $\sigma_{Pt} = 5 \times 10^6 \, \Omega^{-1} \text{m}^{-1}$ [3], and $G^{\uparrow\downarrow} \approx 10^{15} \Omega^{-1} \text{m}^{-2}$ for the Pt-permalloy interface [15-17] together with the assumption that $\text{Re}\, G^{\uparrow\downarrow} >> \text{Im}\, G^{\uparrow\downarrow}$ for typical NM-FM interfaces [14], then with Eqn. (1) the internal spin Hall angle of Pt can be estimated to be $\theta_{SH} \approx 0.20$. This estimate of a quite large $\theta_{SH}$ for Pt as well as the accumulating reports that, depending on the composition and preparation of the Pt-FM bilayer, $\xi_{DL}$ can be quite a bit higher than 0.07 [18-23], point to the opportunity to develop interface engineering protocols that can maximize $\xi_{DL}$ for applications.

Here we report quantitative measurements of the damping-like and field-like torques exerted by the SHE-generated spin currents acting on thin Co and CoFe layers placed in contact with Pt thin film microstrips. We observe variations in the torque efficiencies that depend on the magnetic properties and thickness of the FM layer, and on post-growth thermal processing protocols. We study both perpendicularly-magnetized (PM) structures (Pt-Co, Pt-CoFe) and in-plane magnetized (IPM) structures (Pt-CoFe), measuring the spin-torque efficiencies by the harmonic response (HR) technique [4, 5, 24] (PM cases) and by spin-torque ferromagnetic resonsnce (ST-FMR) [7] (IPM case). Depending on the choice of the FM and the nature of the interface, as characterized by bilayer conductivity and ST-FMR damping coefficient measurements, we find that $\xi_{DL}$ for the Pt-FM system can vary from < 0.05 to > 0.10, with the latter results substantially enhancing the potential value of Pt for three-terminal spin Hall device applications [25].

The multilayer films for this investigation were produced by direct current (DC) sputtering (RF magnetron for the insulating layers) from 2-inch planar magnetron sources onto thermally-oxidized Si substrates in a vacuum system with a base pressure $< 4 \times 10^{-8}$ Torr. We prepared three series of samples: (A) Ta(2)/Pt(4)/Co($t_{Co}$)/MgO(2)/Ta(1) with $0.5\,\text{nm} \leq t_{Co} \leq 1.3\,\text{nm}$, (B) Ta(2)/Pt(4)/Co$_{50}$Fe$_{50}$($t_{CoFe}$)/MgO(2)/Ta(1) with $0.4\,\text{nm} \leq t_{CoFe} \leq 1.1\,\text{nm}$, and (C) Co$_{50}$Fe$_{50}$($t_{CoFe}$)/Pt(4) with $1\,\text{nm} \leq t_{CoFe} \leq 9\,\text{nm}$. The numbers in the parentheses represent the nominal thicknesses of the sputtered films, in nm. The as-deposited series (A) shows perpendicular magnetic anisotropy (PMA) without any further annealing for $0.6\,\text{nm} \leq t_{Co} \leq 1.3\,\text{nm}$. Series (B) requires an annealing temperature of 350°C for 1 hour to obtain PMA, for $0.5\,\text{nm} \leq t_{CoFe} \leq 0.7\,\text{nm}$. Series (C) possesses in-plane magnetic anisotropy for the whole thickness range.

To evaluate $\xi_{DL}$ and $\xi_{FL}$ for these samples we patterned thin film series (A) and (B) into micrometer-sized Hall-bar structures, and measured the HR of the anomalous Hall voltage of these samples to an alternating current passed through the bilayer to determine the effective anti-damping and field-like torques $\tau_{DL}$ and $\tau_{FL}$, or equivalently the longitudinal and transverse effective fields ($H_L \propto \xi_{DL} J_e (\hat{\sigma} \times \hat{m})$ and $H_T \propto \xi_{FL} J_e \hat{\sigma}$) exerted on the perpendicular magnetization. From this and the measured saturation magnetization $M_s$ [26] the spin torque efficiencies, $\xi_{DL}$ and $\xi_{FL}$, were obtained [26]. The IPM series (C) films were patterned into bar-like structures for ST-FMR measurements with $\tau_{DL}$ and $\tau_{FL}$ being determined from the symmetric and anti-symmetric components, respectively, of the ST-FMR signal as discussed below.

In Fig. 1(a) we show both the damping-like and field-like spin-torque efficiencies as determined by HR measurements from the series (A) Pt-Co-MgO samples. The former increases from $\xi_{DL} \approx 0.01$ to $\xi_{DL} \approx 0.11$ as the Co thickness increases from 0.6 nm to 0.9 nm. The latter, although considerably smaller, shows a similar behavior, increasing from $\xi_{FL} \approx 0.00$ for $t_{Co} = 0.6$ nm to $\xi_{FL} \approx 0.03$ at the thick Co limit, $t_{Co} \geq 1.0$ nm. The maximum value for $\xi_{DL}$ that we have obtained here is approximately double, or more, compared to the values reported by early ST-FMR on samples with much thicker permalloy (Py) FM layers ($\geq 4$ nm) [3, 7] and by inverse spin Hall effect (ISHE)-spin pumping measurements of IPM Pt-Py bilayers [27, 28]. This value $\xi_{DL} \approx 0.11$ is also larger than that obtained from PM Pt-CoFe (0.6nm)-MgO structures using the same HR technique ($\xi_{DL} \approx 0.06$) [29] and from magnetic reversal studies on annealed PM Pt-Co-MgO layers formed without a templating Ta seed layer ($\xi_{DL} \approx 0.07$) [30]. However recent ISHE studies of IPM Pt-Py and Pt-YIG have respectively reported $\xi_{DL} \approx 0.12$ [21] and $\xi_{DL} \approx 0.10$ [20], and an even higher value $\xi_{DL} = 0.16$ was found recently for Pt-Co-AlO$_x$ from HR measurements in the PM case [5]. Using ST-FMR measurements, Zhang *et al.* also estimated $\xi_{DL} \approx 0.11$ for Pt-Co bilayer structures [31].

In seeking an explanation for the variation of $\xi_{DL}$ (and $\xi_{FL}$) with $t_{Co}$ it is interesting to note that sputter deposited Ta(seed)-(Pt-Co) multilayers are found to be structurally coherent, with a (111) normal texture in the ultra-thin Co limit, but when the Co becomes thicker than ~ 1 nm the elastic strain due to the ~ 10% lattice mismatch can no longer be supported and the interface relaxes via the formation of misfit dislocations, and the Co and Pt layers become incoherent [32]. Our sputtered Ta(seed)-Pt-Co-MgO heterostructures demonstrate a thickness-dependent magnetoelastic effect associated with interfacial strain (see the Supplementary Material [26] for discussion), which reflects this transition from a strained to a relaxed Co layer. The increase we observe for both $\xi_{DL}$ and $\xi_{FL}$ as a function of increasing $t_{Co}$, therefore, might be related to the transition from a strained and coherent (111) Co layer to a relaxed and incoherent (111) one. Given Eqn. (1), our results also suggest that the increase in the $\xi_{DL(FL)}$ with Co thickness can be associated with an increase in $G^{\uparrow\downarrow}$ as the Pt-Co interface transitions from a coherent (111) to an incoherent (111) interface, although to our knowledge a calculation of $G^{\uparrow\downarrow}$ for such Pt-Co interfaces has not yet been reported.

We performed HR measurements as a function of temperature ($T$) on one of the samples from series (A), namely Ta(2)/Pt(4)/Co(1)/MgO(2)/Ta(1). As shown in Fig. 1(b), no significant $T$-dependence was found for either component of torque down to $T = 50\,\text{K}$. This is strikingly different from recent studies of the Ta-CoFeB-MgO system [33, 34] where a strong $T$-dependence was reported for a much stronger (at $T$ = 300 K) field-like component, together with a $T$ dependence that was weaker, but still significant, for the damping-like torque. Spin Hall magnetoresistance measurements of Pt-YIG bilayer samples [19] have indicated a value $\xi_{DL} \approx$ 0.10 that is approximately independent of $T$ for $T \geq 100\,\text{K}$.

To further examine how the character of the Pt-FM interface can change the behavior of $\xi_{DL}$ and $\xi_{FL}$, we also performed HR measurements on the series (B) samples (Pt-CoFe-MgO, annealed at 350°C to obtain PMA). The resulting $\xi_{DL}$ and $\xi_{FL}$ are shown in Fig. 2(a) as a function of the CoFe thickness $t_{CoFe}$. The first distinct difference compared to the Pt-Co-MgO case is that the magnitude of $\xi_{DL}$ and $\xi_{FL}$ are both high, $\approx 0.15$ for the thinnest layers, $t_{CoFe} = 0.5$ nm, but they *decrease* as a function of increasing $t_{CoFe}$, with $\xi_{FL}$ decreasing more rapidly. A second difference is that the sign of $\xi_{FL}$ is now reversed, negative in our convention. This behavior extends into the IPM regime ($t_{CoFe} \geq 1 \text{nm}$), as discussed below. A third difference is that, as shown in Fig. 2(b), both spin torque efficiencies have a quasi-linear *T*-dependence, with the variation of $\xi_{FL}$ being by far the stronger, so that $\xi_{FL}$ becomes essentially zero by $T = 25$ K while $\xi_{DL} \approx 0.07$ in the low *T* regime. These differences clearly indicate that the nature of spin transport at the annealed Pt-CoFe interface is significantly different from that of the as-deposited Pt-Co interface. Our finding that $\tau_{FL} \approx \tau_{DL}$ in the Pt-CoFe case is of course not immediately consistent with a SHE origin for both torques since calculations find that $\text{Re}\,G^{\uparrow\downarrow} \gg \text{Im}\,G^{\uparrow\downarrow}$ for NM-FM interfaces, even when some disorder is included [14]. However it is also not yet clear how either a Rashba-like effect at the NM-FM or FM-Oxide interface [35, 36], or the SHE, can provide a mechanism to explain the strong *T* dependence of $\tau_{FL}$ for a disordered NM-FM interface.

      To further examine the modifications that can occur in interfacial spin transport due to changes in the details of the Pt-CoFe interface, we carried out ST-FMR measurements on samples of series (C) with thicker CoFe layers, with and without annealing at 350 °C. As discussed in the supplementary materials [26], ST-FMR provides a self-calibrated method for measuring $\xi_{DL}$ *provided* that there is no substantial spin-orbit-induced field-like torque component, $\tau_{FL}$, (equivalent to a transverse effective field) that is comparable to or greater than the Oersted torque $\tau_{Oe}$ from the current in the NM. But if that proviso is not satisfied, then the

self-calibrated ST-FMR analysis is not correct, in that it yields an apparent efficiency, $\xi_{FMR}$, that is not equal to $\xi_{DL}$. Because $\tau_{FL}$ and $\tau_{Oe}$ depend differently on $t_{FM}$, a straightforward way to detect the presence of a significant $\tau_{FL}$ is to perform ST-FMR measurements as a function of FM layer thickness $t_{FM}$; if there is a significant $\tau_{FL}$ the result will be a strong dependence of the apparent torque efficiencies on $t_{NM}$. Previously, a significant $\tau_{FL}$, anti-parallel to $\tau_{Oe}$, has been observed in IPM Ta-CoFeB [37] and Pt-Co [38] bilayer systems as evidenced by a thickness-dependent $\xi_{FMR}$ in the thin $t_{FM}$ limit. For our as-deposited Pt-CoFe bilayers, the self-calibrated ST-FMR analysis yields $\xi_{FMR} \approx 0.10$ with no significant $t_{CoFe}$ dependency for $1\,\text{nm} \leq t_{CoFe} \leq 7\,\text{nm}$ (Fig. 3(a)), indicating that $\tau_{FL}$ is negligible in these as-deposited samples. On the other hand, $\xi_{FMR}$ of the annealed IPM Pt-CoFe bilayers exhibits a sign change at $t_{CoFe} \approx 1.5\,\text{nm}$ (associated with a polarity change of the anti-symmetric Lorentzian signal), which can be explained as due to the presence of a field-like spin-orbit torque and an Oersted torque with opposite signs and changing relative magnitudes. To estimate the spin torque efficiencies in the annealed case, we plot $1/\xi_{FMR}$ against $1/t_{CoFe}^{\text{eff}}$ and perform linear fits for the data (see Fig. 3(b)), where $t_{CoFe}^{\text{eff}}$ represents an effective thickness of CoFe layer excluding the thickness (0.34 nm) of a magnetic dead layer. From the intercept and slope of the fits [26], we estimate $\xi_{DL}^{\text{as-deposited}} = 0.10 \pm 0.005$ and $\xi_{FL}^{\text{as-deposited}} = -0.008 \pm 0.002$, while $\xi_{DL}^{\text{annealed}} = 0.058 \pm 0.006$ and $\xi_{FL}^{\text{annealed}} = -0.028 \pm 0.004$ (Note that this analysis assumes that $\xi_{FL}$ is independent of $t_{CoFe}$ for $t_{CoFe} \geq 1$ nm, which may not be strictly correct.) The substantial changes in $\xi_{DL}$ and $\xi_{FL}$ with annealing demonstrate the sensitivity of the interfacial spin transport to the details of the composition and processing of the NM-FM interface. We note that the X-ray diffraction data for the IPM Pt-CoFe samples before and after annealing indicate a possible transition of the CoFe orientation from strained (111) to relaxed (110) upon annealing, and there is also a significant increase in the CoFe resistance upon annealing [26]. This indicates both a structural change and the creation of greater disorder or perhaps intermixing near the interface, both of which perhaps is affecting the details of the interfacial spin transport, particularly $\text{Im} G^{\uparrow\downarrow}$ which is the measure of spin reflection back from the interface region after some precession within the FM.

As an additional method for examining the interfacial spin transport properties of the IPM Pt-CoFe bilayers, we obtained the effective magnetic damping constant $\alpha$ of the bilayers from the linewidth of the ST-FMR signals for both as-deposited and annealed samples. As shown in Fig. 3(c), while $\alpha$ for both as-deposited and annealed samples increases with decreasing $t_{\text{CoFe}}$, the increase is much more pronounced for the annealed samples, with the thinnest annealed CoFe sample having a damping constant about five times larger than its as-deposited counterpart. According to the theory of spin pumping, the enhancement of $\alpha$ with the placement of a NM next to FM layer can be related to the interfacial effective spin-mixing conductance $g_{\text{eff}}^{\uparrow\downarrow}$ as [39, 40]

$$g_{\text{eff}}^{\uparrow\downarrow} = \frac{4\pi M_s t_{FM}^{\text{eff}}}{\gamma \hbar}(\alpha - \alpha_0) = \frac{4\pi M_s t_{FM}^{\text{eff}}}{\gamma \hbar}\Delta\alpha, \qquad (2)$$

where $M_s$, $t_{FM}^{\text{eff}}$, $\alpha_0$, and $\gamma = 1.76 \times 10^{11}\,\text{s}^{-1}\text{T}^{-1}$ represent the saturation magnetization, the effective thickness, the intrinsic damping constant, and the gyromagnetic ratio of the FM layer, respectively. By plotting $\alpha$ against $1/t_{\text{CoFe}}^{\text{eff}}$, as shown in Fig. 3(d), and performing linear fits based on Eqn. (2), we obtained $g_{\text{eff, as-deposited}}^{\uparrow\downarrow} = (1.5 \pm 0.2) \times 10^{19}\,\text{m}^{-2}$ and $\alpha_0 = 0.008 \pm 0.001$ for our as-deposited Pt-CoFe and $g_{\text{eff, annealed}}^{\uparrow\downarrow} = (12.1 \pm 0.8) \times 10^{19}\,\text{m}^{-2}$ for the annealed Pt-CoFe (the same $\alpha_0$ is employed). While the former result of $g_{\text{eff}}^{\uparrow\downarrow}$ is similar to that reported previously for Pt-FM interfaces [41, 42], the latter is quite large although it is similar to a result recently reported $g_{\text{eff}}^{\uparrow\downarrow} \approx 8.0 \times 10^{19}$ m$^{-2}$ for *as-deposited* Pt-Co bilayers [43]. In the limit of $\text{Re}\,G^{\uparrow\downarrow} \gg \text{Im}\,G^{\uparrow\downarrow}$, the spin-mixing conductance $G^{\uparrow\downarrow}$ can be further calculated from $g_{\text{eff}}^{\uparrow\downarrow}$ by taking the spin back flow at the NM-FM interface into account, with [40]

$$G^{\uparrow\downarrow} = \frac{\frac{\sigma_{NM}}{2\lambda_{s,NM}}\left(\frac{e^2}{h}\right)g_{\text{eff}}^{\uparrow\downarrow}}{\frac{\sigma_{NM}}{2\lambda_{s,NM}} - \left(\frac{e^2}{h}\right)g_{\text{eff}}^{\uparrow\downarrow}\coth\left(\frac{d_{NM}}{\lambda_{s,NM}}\right)}. \qquad (3)$$

Using $d_{\text{Pt}} = 4$ nm, $\lambda_{s,\text{Pt}} \approx 1.4$ nm [3] and the measured Pt conductivity $\sigma_{\text{Pt}} = 3.2 \times 10^6$ $\Omega^{-1}$m$^{-1}$ [26], we find that $G_{\text{as-deposited}}^{\uparrow\downarrow} = (1.28 \pm 0.17) \times 10^{15}$ $\Omega^{-1}$m$^{-2}$. We can then estimate the internal Pt spin Hall angle $\theta_{SH}$ with Eqn. (1) by using this $G_{\text{as-deposited}}^{\uparrow\downarrow}$ and the measured $\xi_{DL}^{\text{as-deposited}}$ (since $|\xi_{DL}| \gg |\xi_{FL}|$ in the as-deposited case this implies that we indeed have $\text{Re}\,G^{\uparrow\downarrow} \gg \text{Im}\,G^{\uparrow\downarrow}$). This yields $\theta_{SH} = 0.21 \pm 0.04$, which is consistent with the estimation from the Pt-Py results and further supporting the finding from the PM Pt-Co measurements that the internal spin Hall angle of Pt is indeed quite large. For the annealed case, if we attempt to apply Eqn. (3) together with the extracted value of $g_{\text{eff}}^{\uparrow\downarrow}$, the result is an unphysical (negative) value for $G^{\uparrow\downarrow}$. We speculate that in the annealed case the increased damping is not due simply to spin-pumping into the Pt (*i.e.*, Eqn. (2)) but rather might be associated with the 0.34 nm magnetic dead layer of CoFe that is formed by the annealing process [26]. If this is the case, the very large value of $g_{\text{eff, annealed}}^{\uparrow\downarrow}$ we extract using Eqn. (2) should be considered unphysical, as well. The influence of the dead layer might also help to explain the relatively large value of $|\xi_{FL}/\xi_{DL}|$ measured for both the annealed PM and the annealed IPM samples.

In summary, we have shown that the efficiency of spin-orbit-induced torques depend quite differently on the FM layer thickness for Pt-Co-MgO and Pt-CoFe-MgO (annealed) PM samples. By optimizing the type and the thickness of FM layer, one can improve the transmission of the spin current from the spin Hall material, Pt in this case, to the adjacent FM layer. Secondly, for in-plane-magnetized Pt-CoFe samples, the efficiencies of the spin-orbit torques vary dramatically as a function of heat treatment and/or different growth protocols such as greater or less energetic sputter deposition or the use of a templating Ta underlayer. This further demonstrates that the structure and quality of the NM-FM interface can play a critical role in spin transport. Most importantly, our results, taken in total with other recent results [18-23], indicate that the "internal" spin Hall angle of Pt is $\theta_{SH} \approx 0.20$. This suggests that more effective and more energy-efficient spintronic devices might be realized by careful Pt-FM interfacial engineering, perhaps surpassing the overall performance of much-more-resistive spin Hall metals having large values of $\theta_{SH}$ (for instance β-Ta [8] and β-W [9]).

C.-F. P. and Y. O. contributed equally to this work.
We would like to thank Graham Rowlands and Junbo Park for the assistance on low temperature measurements, Minh-Hai Nguyen for discussion on the spin-mixing conductance, and Praveen Gowtham for helpful discussion regarding magnetoelastic effects. We thank Stuart Parkin for sharing a manuscript reporting a related study prior to publication. This research was supported in part by NSF/MRSEC (DMR-1120296) through the Cornell Center for Materials Research

[*]rab8@cornell.edu

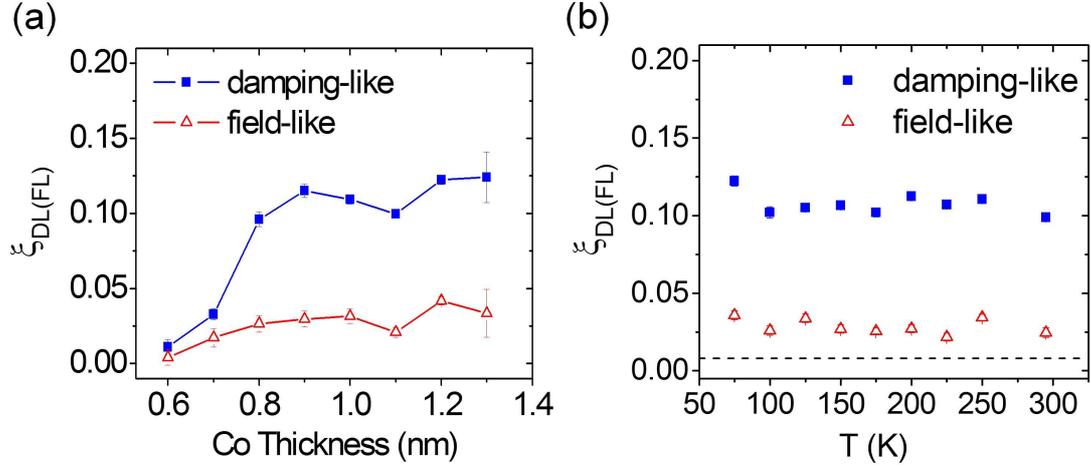

FIG. 1 (color online). (a) The damping-like (field-like) torque efficiency $\xi_{DL(FL)}$ of Pt from as-deposited Pt-Co-MgO samples as a function of Co thickness. Representative error bars due to the device-to-device variation are shown on the data from samples with 1.3 nm of Co. Smaller error bars for other data represent the uncertainties due to fitting. (b) $\xi_{DL(FL)}$ of Pt from a representative Pt(4)/Co(1)/MgO(2) sample as a function of temperature $T$. The blue squares and the red triangles represent data for the damping-like torque and from field-like torque, respectively. For comparison, a dash line is shown in (b) to indicate the sign and the magnitude of the Oersted field expressed in terms of a spin torque efficiency.

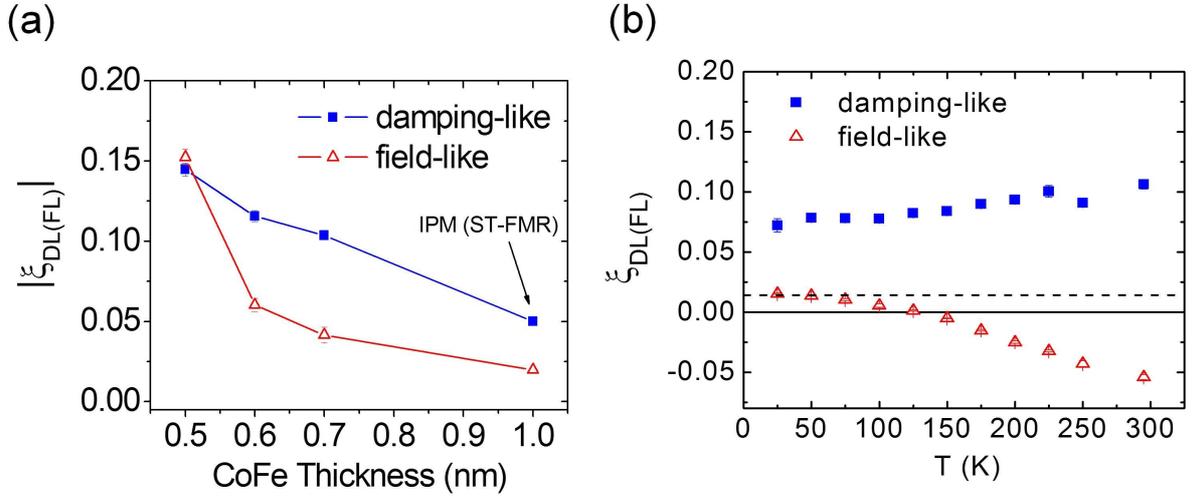

FIG. 2 (color online). (a) The damping-like (field-like) torque efficiency $\xi_{DL(FL)}$ of Pt from Pt-CoFe-MgO samples annealed at 350°C as a function of CoFe thickness. The blue squares and the red triangles represent data for the damping-like torque and the field-like torque, respectively. The data for 1 nm of CoFe were obtained from ST-FMR measurements on IPM samples, while the rest were obtained from HR measurements on PM samples. (b) $\xi_{DL(FL)}$ of Pt from a representative Pt(4)/CoFe(0.6)/MgO(2) sample as a function of temperature $T$. For comparison, a dash line is shown to indicate the sign and the magnitude of the Oersted field in terms of a spin torque efficiency.

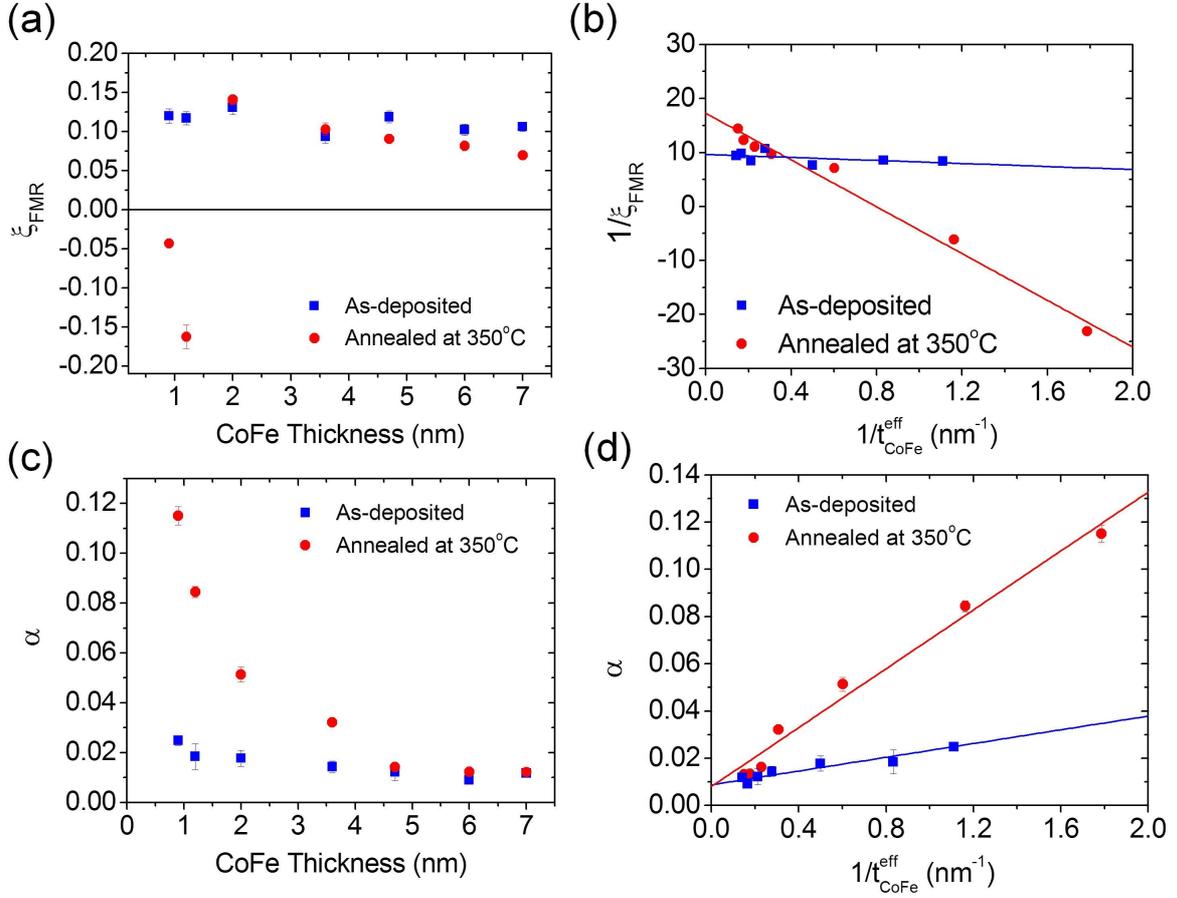

FIG. 3 (color online). (a) The ST-FMR spin torque efficiency $\xi_{FMR}$ of Pt from IPM CoFe-Pt bilayer samples as a function of CoFe thickness by ST-FMR measurements. (b) The inverse of the ST-FMR spin torque efficiency, $1/\xi_{FMR}$, as a function of the inverse of the effective CoFe layer thickness $1/t_{CoFe}^{eff}$. (c,d) The damping constant $\alpha$ of CoFe from as-deposited and annealed CoFe-Pt bilayer structures as a function of (c) CoFe thickness and (d) the inverse of the effective CoFe thickness $1/t_{CoFe}^{eff}$. The blue squares and the red circles represent data from as-deposited and from annealed samples, respectively. The solid lines represent linear fits to the data.

# Supplementary Material

Chi-Feng Pai[1], Yongxi Ou[1], D. C. Ralph[1,2], and R. A. Buhrman[1,*]

[1]Cornell University, Ithaca, New York 14853, USA

[2]Kavli Institute at Cornell, Ithaca, New York 14853, USA

*rab8@cornell.edu

## S1. Saturation magnetization of Co (in Pt-Co-MgO) and CoFe (in Pt-CoFe-MgO)

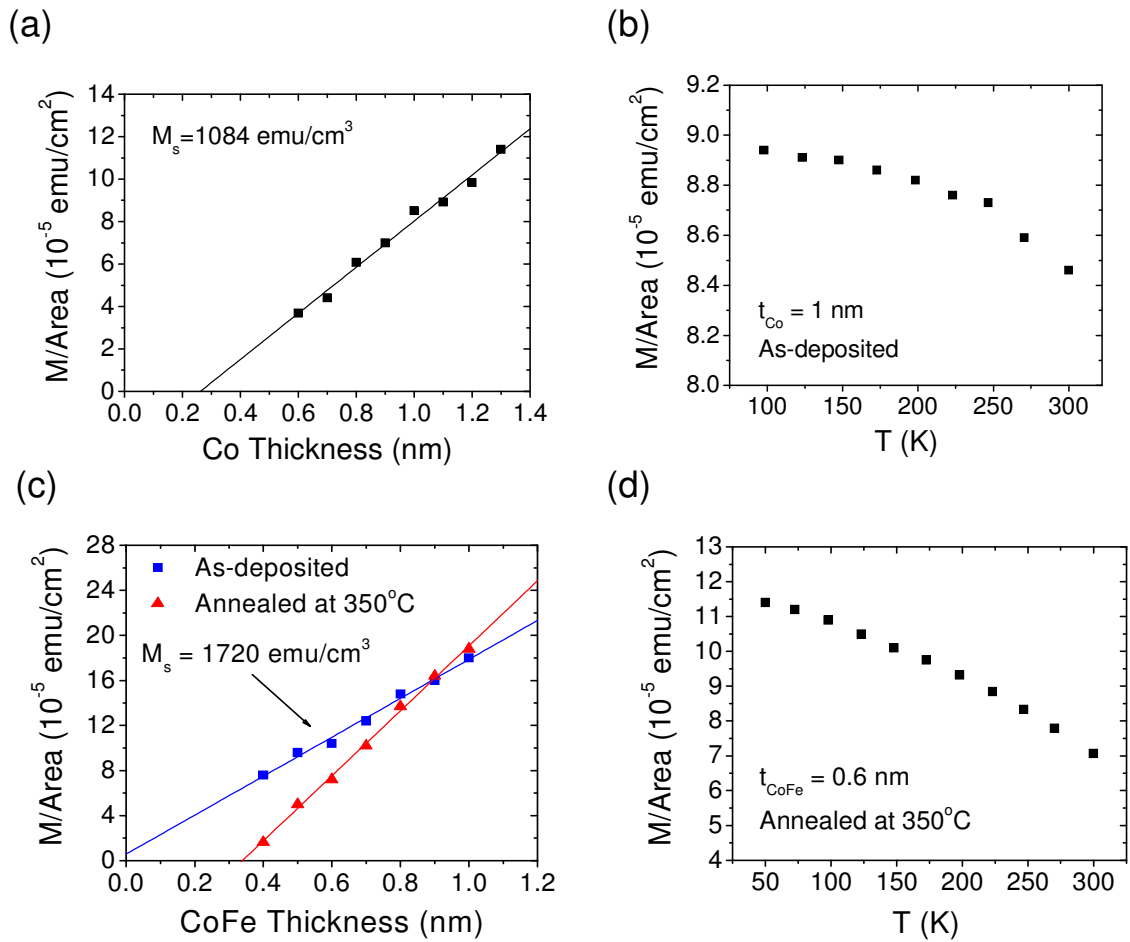

FIG. S1 (color online) (a) The magnetic moment per unit area as a function of sputtered Co thickness in Pt-Co-MgO samples. (b) Temperature dependence of the magnetic moment per unit area in Pt-Co-MgO sample with 1 nm of Co. (c) The magnetic moment per unit area as a function of sputtered CoFe thickness in both as-deposited (blue squares) and annealed (red triangles) Pt-Co-MgO samples. (d) Temperature dependence of the magnetic moment per unit area in annealed Pt-CoFe-MgO sample with 0.6 nm of CoFe.

In order to determine the saturation magnetization $M_s$ of our samples for the estimation of the spin torque efficiencies we measured the FM layer thickness dependence of magnetization for as-deposited Pt-Co-MgO, as-deposited Pt-CoFe-MgO, and annealed Pt-CoFe-MgO samples by superconducting quantum interference device (SQUID) magnetometry. As shown in Fig. S1 (a), the as-deposited Co has a saturation magnetization $M_s = 1084 \pm 50 \, \text{emu/cm}^3$ and a magnetic dead layer thickness of $t_D = 0.26 \pm 0.04 \, \text{nm}$ estimated from the slope and the intercept of the linear fit, respectively. We also measured the temperature dependence of magnetization for $t_{Co} = 1 \, \text{nm}$ sample, which is shown in Fig. S1 (b). Although the magnetization of Co increases while decreasing temperature from 300K to 100K, the variation is less than 10%. For as-deposited CoFe films (See Fig. S1 (c)), we found $M_s = 1720 \pm 70 \, \text{emu/cm}^3$ and no apparent magnetic dead layer. However, a magnetic dead layer of $t_D = 0.34 \pm 0.02 \, \text{nm}$ is formed after the annealing process, with the *apparent* saturation magnetization (from the slope of the linear fit) increases to $M_s = 2884 \pm 60 \, \text{emu/cm}^3$, though the *de facto* total magnetic moment decreases. More importantly, as shown in Fig. S1 (d), the magnetization of our annealed CoFe film with $t_{CoFe} = 0.6 \, \text{nm}$ increases by ~ 60% after cooling from 300K to 50K. Therefore, the temperature $T$ dependence of $M_s(T)$ is an important correction factor, especially for the annealed CoFe case, in order to estimate the effective spin torque efficiency $\xi_{DL(FL)}$ from the measured effective field $H_{L(T)}$ per unit charge current density $J_e$ using (modified from Khvalkovskiy *et al.* [1])

$$\xi_{DL(FL)}(T) = \frac{2e\mu_0 M_s(T) t_{FM}^{eff}}{\hbar}\left(\frac{H_{L(T)}}{J_e}\right), \quad (S.1)$$

where $t_{FM}^{eff} = t_{FM} - t_D$ represents the effective thickness of the FM film. All the spin torque efficiencies $\xi_{DL(FL)}$ shown in the main text were calculated with the correction from the temperature dependency of $M_s$ and the existence of the dead layer.

## S2. Magnetic anisotropy energy in Pt-Co-MgO and in Pt-CoFe-MgO heterostructures

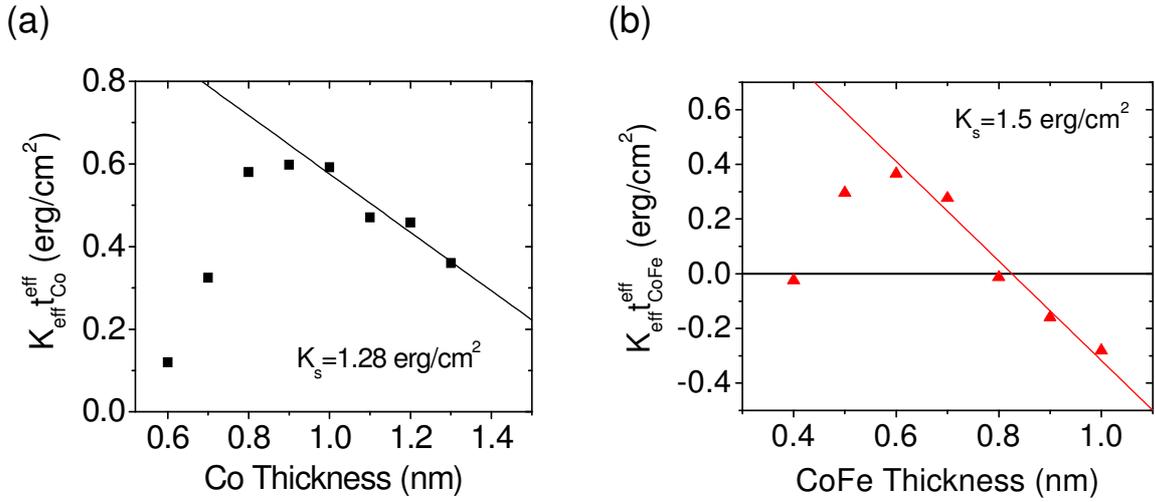

FIG. S2 (color online) The effective magnetic anisotropy energy density in terms of $K_{eff} t_{FM}^{eff}$ (a) for as-deposited Pt-Co-MgO as a function of Co thickness and (b) for annealed Pt-CoFe-MgO as a function of CoFe thickness. $K_s$ indicates the interfacial anisotropy energy obtained from the intercept of the linear fit (solid lines).

We characterized the magnetic anisotropy energy density $K_{eff}$ in both as-deposited Pt-Co-MgO and annealed Pt-CoFe-MgO samples by SQUID magnetometry (to obtain $M_s$) and anomalous Hall voltage measurement (to obtain the anisotropy field $H_{an}$ from our samples). The results are summarized in Fig. S2. We performed linear fits [2] to the data of $K_{eff} t_{FM}^{eff}$ from the linear regime, which corresponds to the regime of the relaxation of lattice strain in the FM layer [3], to estimate the interfacial anisotropy energy $K_s$ from the intercept for both cases ($K_s = 1.28 \pm 0.17 \, \text{erg/cm}^2$ for Pt-Co-MgO and $K_s = 1.50 \pm 0.24 \, \text{erg/cm}^2$ for annealed Pt-CoFe-MgO). From the trend of $K_{eff} t_{FM}^{eff}$ curve, we also speculate that the Co thickness range that gives rise to a $\xi_{DL} \approx 0.11$ $\xi_{FL}$ in Pt-Co-MgO samples corresponds to the regime that the strain has been relaxed and an incoherent Pt-Co (111) interface is formed. However, for the annealed Pt-CoFe-MgO case, the CoFe thickness that shows $\xi_{DL} \approx \xi_{FL} \approx 0.15$ corresponds to the coherent regime. Perhaps the decrease in $\xi_{DL}$ and $\xi_{FL}$ with increasing $t_{CoFe}$ can be attributed to relaxation via misfit dislocation of the elastic strain in the CoFe, or some other change in the layer. (See S4 below.)

## S3. ST-FMR signals and analysis from IPM Pt-CoFe samples

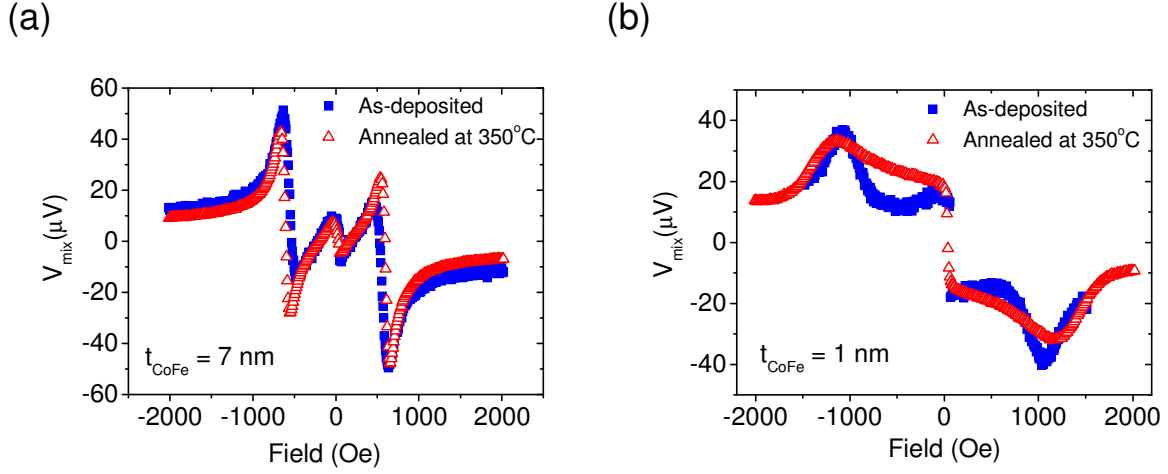

FIG. S3 (color online) The representative ST-FMR $V_{mix}$ signals from (a) as-deposited and annealed CoFe(7)/Pt(4) samples and from (b) as-deposited and annealed CoFe(1)/Pt(4) samples. The blue solid squares and the red open triangles represent data from as-deposited and from annealed samples, respectively.

The representative ST-FMR mixing voltage ($V_{mix}$) signals [4] for as-deposited and for annealed CoFe-Pt bilayer samples are shown in Fig. S3 (a) for thick CoFe regime and (b) for thin CoFe regime. In the thin CoFe regime, the effective field originating from the field-like torque is anti-parallel to the Oersted field, which leads to a sign change in the anti-symmetric Lorentzian component of the $V_{mix}$ signal. By taking the field-like torque into account, we can modify the formula given by Liu *et al.* [4] into

$$\xi_{FMR} = \frac{S}{A'} \frac{e\mu_0 M_s t_{FM}^{eff} d_{NM}}{\hbar} \sqrt{1+\left(4\pi M_{eff}/H_0\right)}. \quad (S.2)$$

The new $S$ and $A'$ now respectively stand for

$$S = \frac{\hbar \xi_{DL} J_e^{rf}}{2e\mu_0 M_s t_{FM}^{\text{eff}}} \quad \text{and} \tag{S.3}$$

$$A' = \left(H_T + H_{Oe}\right)\sqrt{1+\left(4\pi M_{\text{eff}}/H_0\right)} = \left(\frac{\hbar \xi_{FL} J_e^{rf}}{2e\mu_0 M_s t_{FM}^{\text{eff}}} + \frac{J_e^{rf} d_{NM}}{2}\right)\sqrt{1+\left(4\pi M_{\text{eff}}/H_0\right)}, \tag{S.4}$$

where $t_{FM}^{\text{eff}}$, $d_{NM}$, $4\pi M_{\text{eff}}$, $H_0$, $J_e^{rf}$, $H_T$, and $H_{Oe}$ represent the effective thickness of FM layer, the thickness of normal metal (NM) layer, the effective demagnetization field of the FM layer, the resonance field, the rf charge current density in the NM layer, the rf effective field (from the rf current-induced field-like torque $\tau_{FL}$) and the rf Oersted field, respectively. Some math using the above equations will give us the fitting function that we employed in the main text to extract the "effective" spin torque efficiency for the damping-like and field-like torques,

$$\frac{1}{\xi_{FMR}} = \frac{1}{\xi_{DL}}\left(1 + \frac{\hbar \xi_{FL}}{e\mu_0 M_s t_{FM}^{\text{eff}} d_{NM}}\right). \tag{S.5}$$

**S4. Effective magnetic anisotropy energy density and conductance of IPM Pt-CoFe samples**

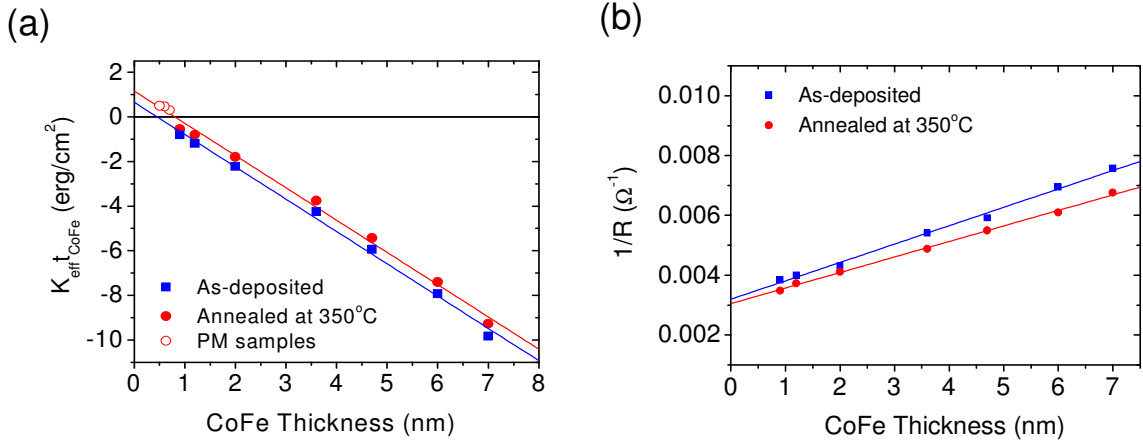

FIG. S4 (color online) (a) The effective magnetic anisotropy energy density in terms of $K_{\text{eff}} t_{\text{CoFe}}$

and (b) the electrical conductance against the sputtered CoFe thickness. The solid blue squares and the red circles represent data from as-deposited and annealed samples, respectively. The open circles in (a) represent data from annealed PM Pt-CoFe-MgO samples.

We also characterized the effective magnetic anisotropy energy density $K_{eff}$ (from unpatterned thin films) and the conductance $1/R$ (from bar-like devices with lateral dimensions $5\mu m \times 20\mu m$) for as-deposited and annealed IPM Pt-CoFe bilayer samples. As shown in Fig. S4 (a), the interfacial anisotropy energy density of IPM Pt-CoFe films, which was estimated from the intercept of the linear fit [2], is increased from $K_s^{\text{as-deposited}} = 0.65 \pm 0.18 \, \text{erg/cm}^2$ to $K_s^{\text{annealed}} = 1.16 \pm 0.20 \, \text{erg/cm}^2$. For reference, the results from the Pt-CoFe-MgO (annealed) with perpendicular magnetic anisotropy are also plotted on the same figure, which indicates a small enhancement of $K_s$ with the addition of the CoFe-MgO interface. The resistivities of the CoFe layer and the Pt layer, which were estimated from the conductance data shown in Fig. S4 (b), also demonstrate slight variations after the annealing process. The resistivity of Pt changes from $\rho_{Pt}^{\text{as-grown}} = 31.3 \, \mu\Omega \times \text{cm}$ to $\rho_{Pt}^{\text{annealed}} = 32.8 \, \mu\Omega \times \text{cm}$, while the resistivity of CoFe changes from $\rho_{CoFe}^{\text{as-grown}} = 40.7 \, \mu\Omega \times \text{cm}$ to $\rho_{CoFe}^{\text{annealed}} = 48.2 \, \mu\Omega \times \text{cm}$ which suggests increased disorder in the CoFe or some diffusion of Pt into the CoFe near the interface. These variations indicate that both the magnetic and the transport properties of the Pt-CoFe bilayer structure are significantly modified after the heat treatment.

**S5. X-ray diffraction patterns of Pt-CoFe samples before and after annealing**

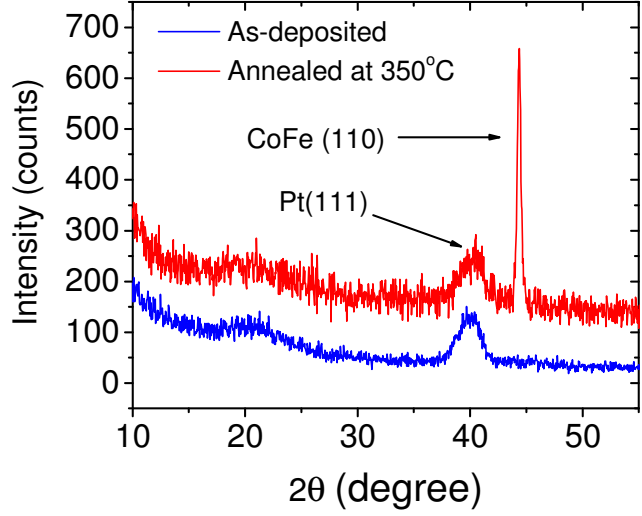

FIG. S5 (color online) X-ray diffraction patterns ($2\theta-\omega$ scan) for the as-deposited (the lower trace line) and the annealed (upper trace line) CoFe(3)/Pt(4) bilayer samples.

We further performed X-ray diffraction ($2\theta-\omega$ scan) measurements on both as-deposited and annealed CoFe(3)/Pt(4) IPM films to study possible structural change in these bilayer samples. As shown in Fig. S5, the 4 nm Pt and the 3 nm CoFe together show a broad (111) peak (similar to the case of Pt-Co multilayer as shown in Ref.[3]) for the as-deposited sample. After annealing, we found a sharp CoFe (110) peak [5, 6], which indicates that the high temperature annealing process promotes the re-crystallization of the thin CoFe layer. This is consistent with our obtaining perpendicular magnetic anisotropy in the Pt-CoFe-MgO samples *only* after the heat treatment, since the CoFe therein will then re-orient with respect to the MgO (100) layer [6]. The reduction of the damping-like spin torque efficiency $\xi_{DL}$ from 0.10 to 0.05 after annealing for the IPM Pt-CoFe case, therefore, can be possibly attributed to this dramatic structural change of CoFe from (111) to (110) orientation.